\documentclass[journal,a4paper,twoside,10pt]{IEEEtran}

\usepackage{amsmath}
\usepackage{cite}
\usepackage{graphicx}
\usepackage{booktabs}
\usepackage{amsfonts}
\usepackage{epsfig}
\usepackage{psfrag}
\usepackage{latexsym}
\usepackage{footnote}
\graphicspath{ {pictures/} }

\usepackage[usenames, dvipsnames]{color}
\usepackage{xcolor}

\begin{document}

\title{A Deep Autoencoder System for Differentiation of Cancer Types Based on DNA Methylation State}

\author{Mohammed~Khwaja,~\IEEEmembership{Student Member,~IEEE,} Melpomeni~Kalofonou,~\IEEEmembership{Member,~IEEE} and~Chris~Toumazou,~\IEEEmembership{Fellow,~IEEE}
\thanks{Authors are with the Department of Electrical and Electronic Engineering, Imperial College London, SW7 2BT, United Kingdom. Email: \{mohammed.khwaja16, m.kalofonou, c.toumazou\}@imperial.ac.uk.}
\thanks{M. Kalofonou and C. Toumazou are also with the Centre for Bio-Inspired Technology, Institute of Biomedical Engineering, Imperial College London, SW7 2BT, United Kingdom.}}

\maketitle

\vspace{-2cm} 

\begin{abstract}

A Deep Autoencoder based content retrieval algorithm is proposed for prediction and differentiation of cancer types based on the presence of epigenetic patterns of DNA methylation identified in genetic regions known as CpG islands. The developed deep learning system uses a CpG island state classification sub-system to complete sets of missing/incomplete island data in given human cell lines, and is then pipelined with an intricate set of statistical and signal processing methods to accurately predict the presence of cancer and further differentiate the type and cell of origin in the event of a positive result. The proposed system was trained with previously reported data derived from four case groups of cancer cell lines, achieving overall Sensitivity of 88.24\%, Specificity of 83.33\%, Accuracy of 84.75\% and Matthews Correlation Coefficient of 0.687. The ability to predict and differentiate cancer types using epigenetic events as the identifying patterns was demonstrated in previously reported data sets from breast, lung, lymphoblastic leukemia and urological cancer cell lines, allowing the pipelined system to be robust and adjustable to other cancer cell lines or epigenetic events.

\end{abstract}


\IEEEpeerreviewmaketitle
\section{Introduction}

Significant progress has been made in understanding crucial regulatory mechanisms responsible for the development and progression of cancer at a cellular and molecular level, through genetic alterations such as DNA mutations and disruptions in epigenetic mechanisms including DNA methylation and histone modifications~\cite{esteller2008epigenetics}. Cancer rates have been progressively increasing, with the latest statistics from Cancer Research UK to have reported more than 350,000 new cases diagnosed in the UK~\cite{cancer2014uk}, of which more than 40\% could have been prevented. Cancer research has been significantly progressing with advances in more effective treatments and screening methods, however there is still a pressing need for more targeted methods to be available for monitoring of cancer progression and prevention of treatment resistance that would help control the disease and improve survival rates. With this goal, cancer diagnostics have been evolving with findings from ongoing clinical trials providing strong evidence on the importance and translational significance of cancer specific biomarkers aiming towards more precise cancer prediction systems and monitoring schemes~\cite{shaw2017mutation}. 

Given that cancer mortality rates are higher for patients diagnosed with advanced staged disease, early diagnosis of progressive disease and personalisation of treatment is essential to achieve better response and therefore higher chances of prognosis. The use of well-defined epigenetic biomarkers of predictive and prognostic value, such as DNA methylation, is essential to be considered in the panel of markers for monitoring of cancer development and treatment efficacy, due to their role in critical pathways and regulatory mechanisms of gene expression~\cite{dawson2012cancer, flanagan2015epigenome}.
 
\begin{figure}[!t]
\centering
\includegraphics[width=7 cm]{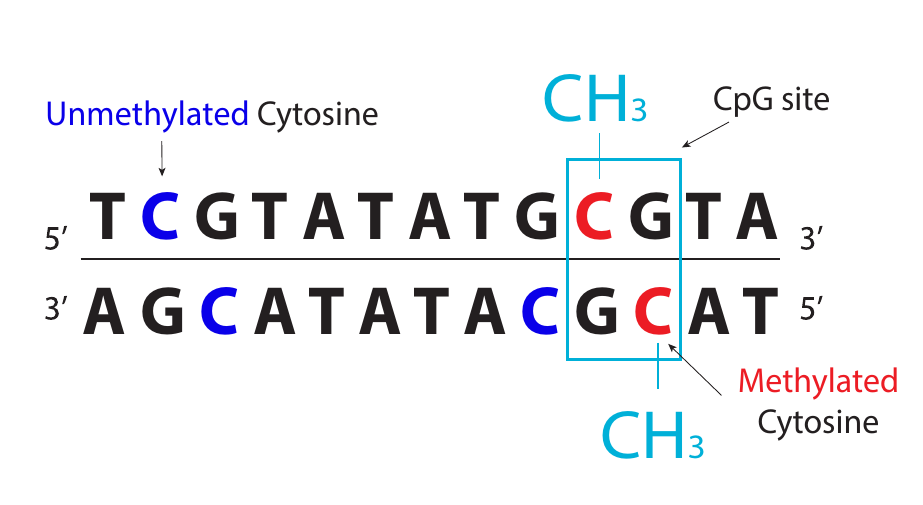}
\caption{Representation of a genetic region with a CpG site and methylated/unmethylated cytosines indicated in red and blue respectively.}
\label{fig:cpg_methylation}
\vspace{-1em}
\end{figure}

Studies have indicated that aberrations in regulatory areas of tumour suppressor genes or oncogenes can be caused by disruptions in DNA methylation patterns, in genome-wide regions as well as genetic regions with high concentration of C and G bases, also known as CpG islands, leading to 'gene silencing', affecting critical switching mechanisms related to DNA repair and tumour suppression~\cite{doi2009differential}. Most CpG islands are located within the 5' region of expressed genes although they can also be found near the 3' region, with DNA methylation to occur most commonly at the cytosine base of DNA, defined by the addition of a methyl group ($CH_3$) to the C-5 position, as shown in Fig.~\ref{fig:cpg_methylation}. Therefore, in combination with genetic mutations, detection of DNA methylation aberrations is of great importance for the development of future generation of closed-loop cancer monitoring systems, combining sensing modalities through Lab-on-CMOS systems in combination with data analytics and classification algorithms~\cite{kalofonou2012isfet, kalofonou2014isfet, khwaja2017}.

While CMOS based sensor technologies have been proven very effective for diagnostic applications given the inherent advantages of speed, scalability, low cost and the capability to integrate sensing and processing units in System-on-Chip interfaces~\cite{toumazounature, moser2018scalable, moser2016isfets, kalofonou2014low}, the growing requirements for processing of large data sets has created a unique opportunity for machine learning algorithms to be incorporated into existing systems given the requirements for real-time analytics and processing of large data sets. For cancer diagnostics in particular, such synergy is foreseen to allow for the development of more precise, adaptable, intelligent and self-learning systems, assisting diagnosis and prediction of cancer risk through the advancement of bioinformatics frameworks, reducing the inherent complexities in analysis of large-scale data. Recently, numerous methods have started to explore the use of DNA methylation patterns analysed from thousands of cancer genes, using classical and advanced machine learning algorithms, to create systems that can predict and differentiate cancer types~\cite{si2016learning, kang2017cancerlocator, lehmann2016identification, hao2017dna}. These state-of-the-art methods present unique ways of exploring methylation patterns, but miss out on using core CpG island attributes including the occurrence of dinucleotides, DNA energies of octamers and other distinct features to provide rich, additional information for resilient learning of CpG differences. 

In this paper, a deep learning system was developed for prediction and classification of DNA methylation patterns, used as the distinguishing factors between healthy and of cancerous origin data sets. The system incorporates the use of an algorithm for identifying and completing missing CpG island states and then processes the extracted island information, identifying essential features required for the differentiation of data using well-defined mathematical algorithms. The ability to learn and autonomously distinguish various cancer types has been achieved through the use of a Deep Autoencoder and the combination of an island state prediction algorithm and of statistical methods. The efficiency of the system was demonstrated for four cancer types, showing that the proposed method can selectively isolate relevant features and differentiate between each other, ensuring robustness of the system and adaptability in different data sets.
 
\section{Machine Learning for tumour classification}
\label{sec:previous}

Tumor classification based on machine learning algorithms using DNA methylation as a recognition pattern has been a fast growing area of research, with recently reported work predominantly based on the use of learning models and dimensionality reduction methods to locate and then predict the presence of DNA methylation in given data sets. Specifically, Zhongwei et al.~\cite{si2016learning} used high dimensional DNA methylation data, obtained from the Illumina 27k array to predict the presence or absence of breast cancer in a given cell type. The learning model was based on identifying CpG island methylation points in cell lines as the feature set, through a stack of Random Boltzmann Machine (RBM) layers to produce a deep learning structure, while reducing dimensionality of the feature set. The model first selected the best 5,000 features based on variance from a set of over 27,000 features and subsequently used four RBM layers to reduce the number of features to 10-70, followed by classification of generated features, suggesting that using a stack of RBMs combined with Self Organised Maps (SOM) to cluster, achieved better feature selection compared to more classical methods including Principal Component Analysis (PCA), K-means and Gaussian mixture model (GMM), with an overall prediction error rate of 2.94\% extracted from one database for a single cancer binary classification. 
Followingly, CancerLocator was proposed as a prediction method for the presence and location of tumours using a binomial probabilistic model, creating a link between circulating cell-free DNA (cfDNA) in blood samples and genome wide DNA methylation~\cite{kang2017cancerlocator}. Although recent methods have proposed multi-cancer classification methods~\cite{lehmann2016identification}, the CancerLocator model demonstrated good results for different cancer types in real and simulated data, as compared to Random Forest (RF) and Support Vector Machine (SVM). Concurrent work on DNA methylation for diagnosis of cancer types was also demonstrated by Hao et al.~\cite{hao2017dna}, using the Least Absolute Shrinkage and Selection Operator (LASSO) multinomial distribution, achieving classification of four cancer types with high accuracy.

Previously reported algorithms have not taken however into account the effect of missing data in CpG island features for DNA methylation prediction. Recently proposed deep learning methods have been illustrated to determine missing CpG island states, including examples by~\cite{wang2016predicting, khwaja2017}. These, in combination with highly advanced feature-selection algorithms using deep learning systems, could provide substantial improvements on performance characteristics and accuracy of data prediction. The proposed system presented here is based on a previously developed DNA methylation prediction algorithm for classification of DNA methylation states using a Deep Belief Network~\cite{khwaja2017}, introducing as a new feature the novel characteristic of using a Deep Autoencoder based deep learning content retrieval algorithm to predict the presence of cancer types in previously confirmed cancer cases, with an additional feature of classification of DNA methylation values as the differentiating factor. Through this approach it is therefore offering enhanced accuracy of prediction by completing missing or incomplete data in the data sets, which tend to be typically the origin for accuracy variation.

\section{Under study Data-sets}
\label{sec:data-set}

Four categories of cancer types and cell lines in particular were used in this study to demonstrate efficacy of the proposed method, obtained from a number of previously reported data-sets:

\begin{itemize}

\item \textit{Breast cancer}: With a lifetime incidence risk of 12\%, and 20-30\% of diagnosed and treated patients to relapse in the form of an aggressive metastatic disease, breast cancer is one of the most common cancer types, raising even more the importance to be studied and classified. Obtained from~\cite{heyn2012dna}, data from 30 samples were used, equally divided in 2 pairs of 15 representing healthy and cancerous cells respectively. 

\item \textit{Lung carcinoma}: Being one of the most frequently detected cancer types, lung cancer is also linked to considerably low survival rates. Obtained from~\cite{karlsson2014genome}, a database with an extensive set of five lung cancer subtypes was used, including 124 tumour and 12 healthy samples.

\item \textit{Lymphoblastic leukemia}: Procured from~\cite{busche2013integration}, this data-set provides samples from B-cell precursor acute lymphoblastic leukemia (pre-B ALL), one of the most common types of pediatric cancer \cite{cheok2006acute}. 92 data samples were used, 46 of which were derived from tumor samples and 46 from remission samples.

\item \textit{Urological tumors}: Attained from~\cite{ramalho2017downregulation}, this data-set comprises of data sets from 16 healthy and 67 cancerous samples. The data-set is further divided into three subtypes: kidney with 6 healthy and 17 cancer cells; bladder with 5 healthy and 25 cancer cells; and prostate with 5 healthy and 25 cancer cells.

\end{itemize}

In total, 341 samples were used to test the efficacy of the proposed system, of which 252 were tumor derived samples with the rest being cancer free, with specific CpG methylation profiles of human cells extracted from the Illumina Human450 BeadChip platform~\cite{bibikova2011high}. After multiple tests with various ratios being performed, 75\% of the total number of samples were selected to be used for training, 9\% for validation and 16\% for testing, with the selection of samples percentages to have been a result of optimisation during testing while avoiding overfitting to occur.

Each of the samples in the first three data-sets contained approximately 485,577 CpG islands, with a small variation in the urological tumor data sets that contained 485,512 CpG islands. Random sample selection was performed 10 times and the results were averaged with tests results derived from a K-fold validation. In each specimen of every category, a number of CpG islands was missing as data may have been lost, incorrectly detected or processed. In previously reported methods for distinguishing cancer types, in the event of a missing data set derived from an island occuring, data from the same island from all samples were consequently discarded, causing loss of essential features which could have otherwise provided useful information to the learning network. This challenge was tackled with the proposed solution, with the following sections to present the methods and resulting performance of the proposed system, overcoming the issue of incomplete data bypassing.

\section{Proposed Deep Learning System}
\label{sec:method}

\begin{figure}[!t]
\centering
\includegraphics[width=\linewidth]{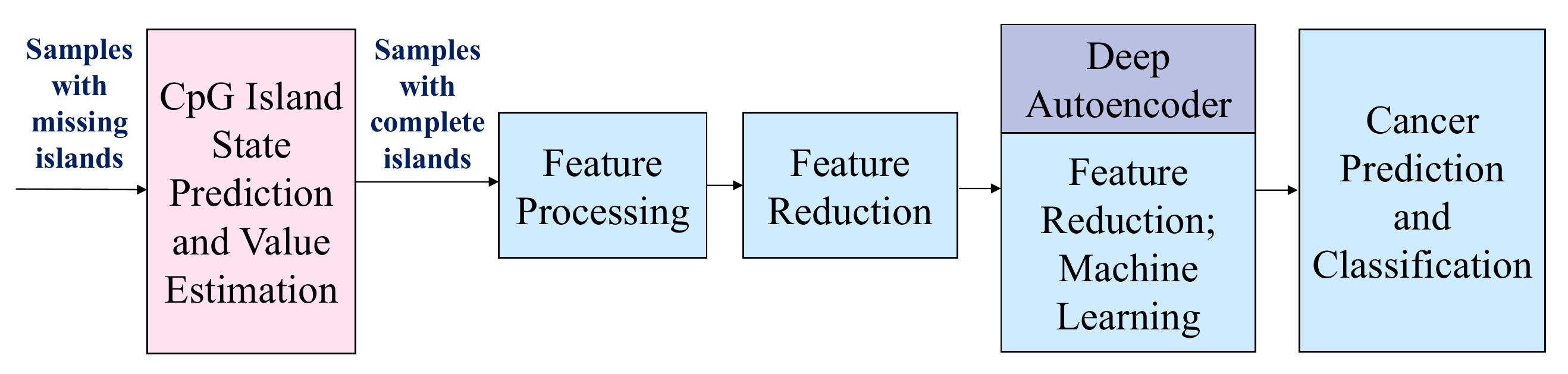}
\caption{Proposed Deep Autoencoder framework incorporating the DNA methylation state prediction system.}
\label{fig:FlowChart_Disease}
\end{figure}

The proposed cancer prediction and classification algorithm is illustrated in Fig.~\ref{fig:FlowChart_Disease}. It first uses the previously reported methylation prediction method ~\cite{khwaja2017} to determine the state of missing/incomplete CpG islands, providing comprehensive information derived from given cell lines. As each cell line possesses large features (485,577 islands), the collection of all samples can be considered as equivalent to a big data set. As with any big data structure, a series of intricate and connected feature extraction and reduction methods are needed to reduce computational complexity, reduce time and increase efficiency of data distribution. The island descriptors are processed and reduced by a set of methods described in Section~\ref{sec:feature}, and once a reduced state is obtained, a Deep Autoencoder network is implemented, as described in Section~\ref{sec:dae}, to perform further dimension reduction and learning. Subsequent dimension reduction using a Deep Autoencoder is cemented to the learning network by an important factor, providing the ability to improve its own efficiency in predicting the presence of cancer and classifying its type.

\subsection{Prediction of missing CpG methylation data}
\label{sec:missing}

All missing islands, incorrectly obtained data and incomplete data-sets are completed using the methylation prediction algorithm described in~\cite{khwaja2017}, to provide comprehensive information to all data sets, so that all cell lines contain uniform and complementary information.

\begin{figure}[!t]
\centering
\includegraphics[width=\linewidth]{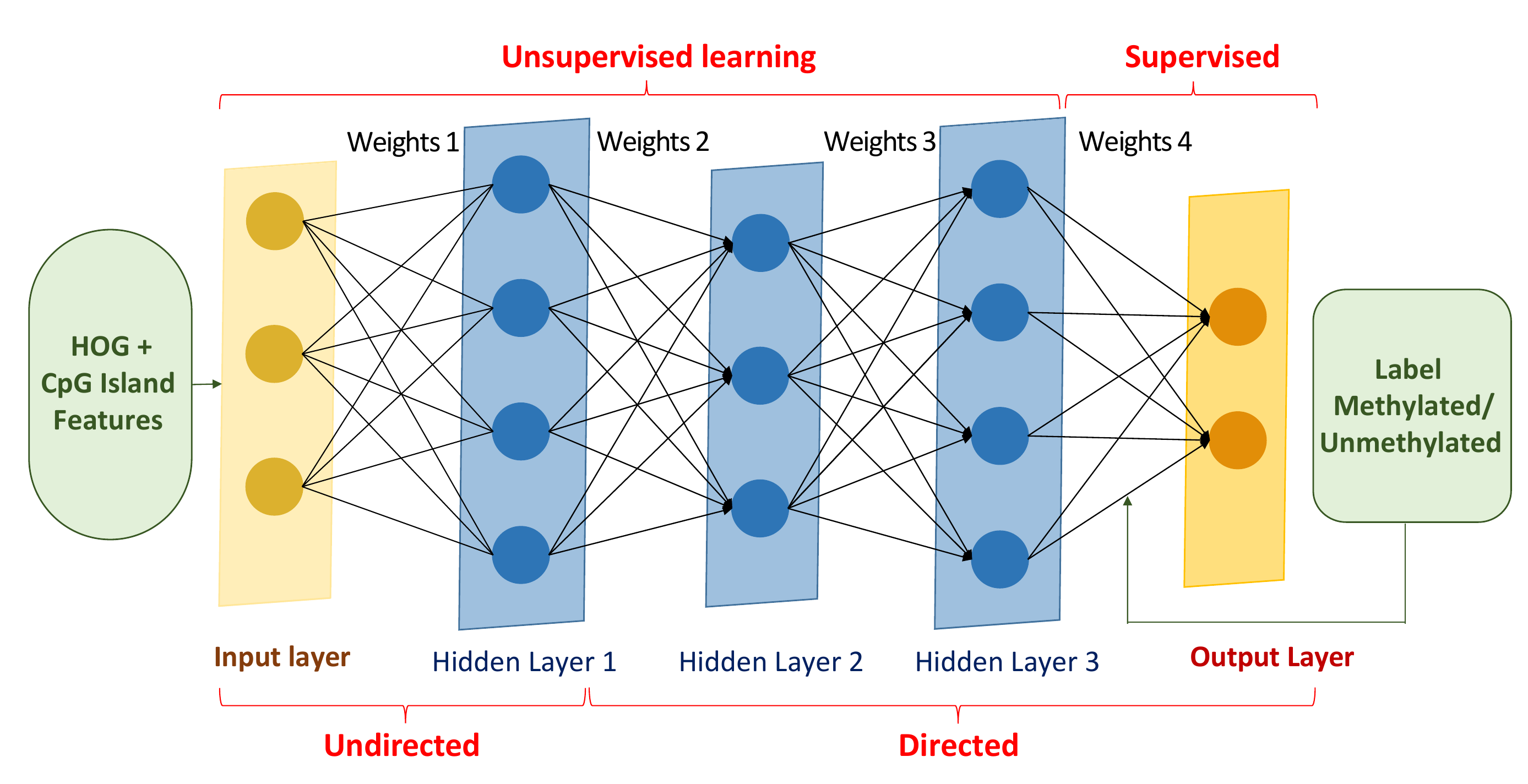}
\caption{Deep Belief Network (DBN) structure used for the prediction of DNA methylation state.}
\label{fig:dbn_layers}
\end{figure}

The algorithm described in~\cite{khwaja2017} obtained methylation characteristic features for every CpG island and included parallel image-analogous islands descriptors to derive the histogram of oriented gradients (HOG) features, which were then used as inputs of the Deep Belief Network (DBN) to classify islands as either methylated or unmethylated. A representation of the DBN structure can be seen in Fig.~\ref{fig:dbn_layers}. This algorithm is used as it is highly accurate and consistent, and does not depend on the area of the genome or chromosome under test. Training of the island classification system is then performed using data from the ENCODE consortium~\cite{encode2012integrated}, while classification is achieved for the data-set described in Section~\ref{sec:data-set}. 

Prior to the classification, the average value of all hypermethylated and hypomethylated islands of all cell lines are separately obtained. All islands that possess a methylation value greater than 0.5 are considered as 'hypermethylated' and those below 0.5 as 'hypomethylated'. For islands classified as methylated, the average value of a hypermethylated state is substituted, while for those classified as unmethylated, the average value is hypomethylated. Thus, each cell line now has comprehensive information for all 485,577 islands through normalisation of all data sets, preventing potential loss of essential data that used to be eliminated by previously reported methods, while allowing for detection and differentiation of various cancer types.

\subsection{Observation of features and preliminary interpretations}
\label{sec:obs}

\begin{figure*}
\centering
\includegraphics[width=\linewidth]{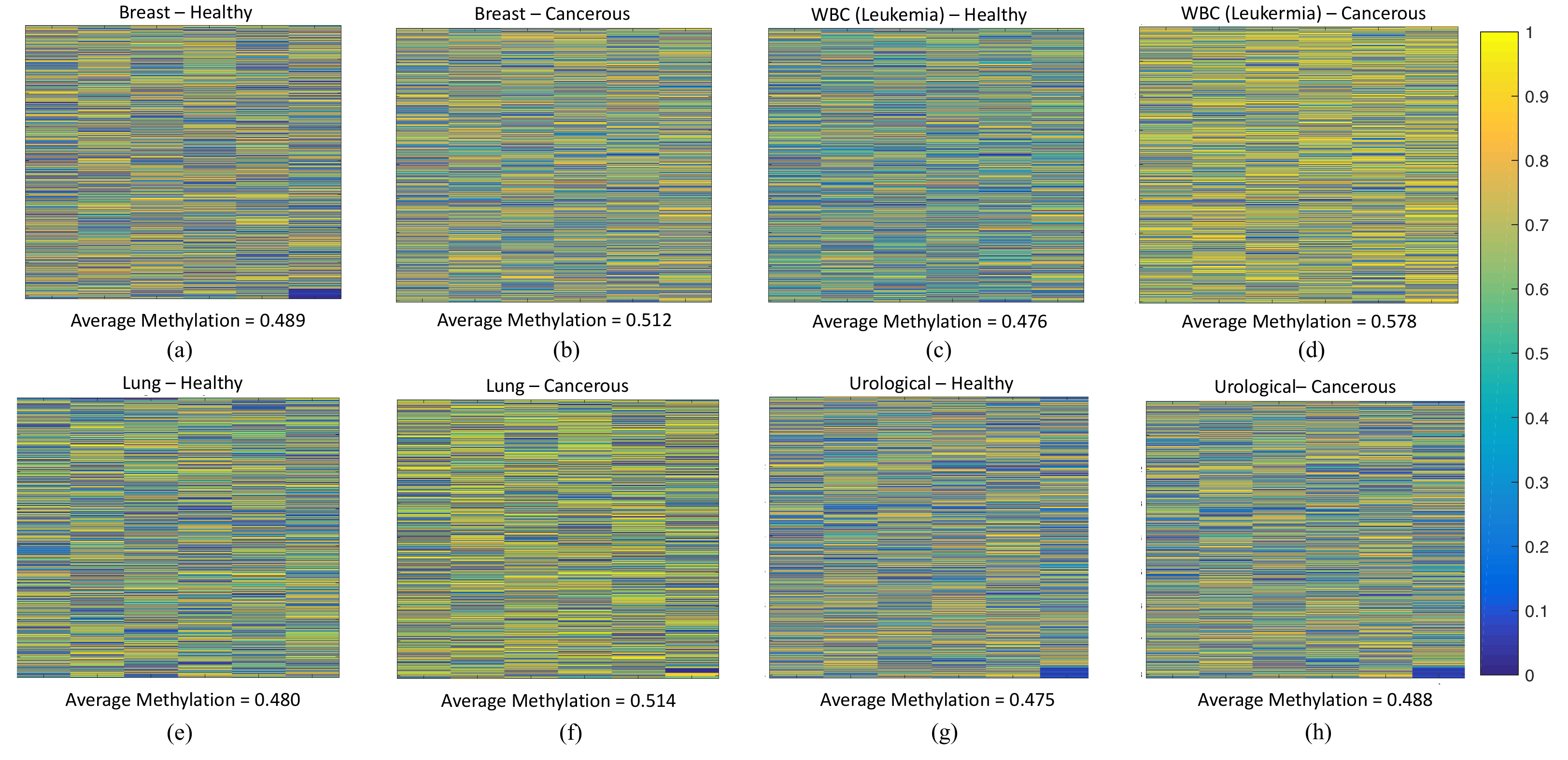}
\caption{Heatmaps of healthy cells (odd figures) and cancerous cells (even figures) for each cell type. The cells were randomly selected from the data-set. The colour spectrum depicts high intensity values appearing in yellow, and low ones appearing in blue. Healthy cells from breast, leukemia and lung samples possess large areas of green (median of colour spectrum), indicating average methylation, while their cancerous counterparts have high intensity values interspersed with low values (combination of extremes in the spectrum); the healthy and cancerous cells chosen from the urological samples possess mostly median range values with small amounts of extremes.}
\label{fig:Heatmaps}
\vspace{-0.5em}
\end{figure*}

Before performing feature transforms, the cell lines are visually analyzed. On the basis that cancer cells possess local increase (hypermethylation) but global reduction (hypomethylation) of DNA methylation~\cite{esteller2008epigenetics}, after collecting comprehensive information for all cell lines, heatmaps for various cancerous and healthy cells were obtained and plotted in Fig.~\ref{fig:Heatmaps}, as a method of visually illustrating the increase/decrease of methylation. It is evident that global and local differences appear in the cases of breast, lung and leukemia cells. Interestingly, when all cases are viewed as a complete image, cancer cells global methylation intensity appears to be high even though the average values are not. This is attributed to the fact that cancer cells have large areas of hypermethylation (regions appearing in yellow) locally dispersed with regions of hypomethylation (regions appearing in blue), causing the average to balance out. Healthy cells tend to maintain an average value in methylation across all regions (seen by the large green regions). This concurs with the expectation previously stated. Of the above three cell lines, the difference between leukemia's cancer and healthy cell lines is noticeably visible and leads to the observation that classification of leukemia samples is expected to be highly accurate. This trend is not followed as clearly in the case of urological samples, as the cancerous and healthy cells seem to possess similar heatmaps leading to identical feature sets. Thus, it is expected that the classification of urological samples may not achieve the desired results for the given database. The heatmaps provide useful information and a visual representation of the data-sets, and to accomplish robust learning of all the data-sets, the next section describes the extraction, processing and selection of high calibre features needed to feed the Autoencoder system. 

\subsection{Feature processing and reduction}
\label{sec:feature}

For a neural network to form a good mapping distribution of features, it is essential that quality descriptors are used by transforming features into standard and normalized values so that each neuron in the network creates a good weighted response to its inputs. Feature selection and extraction (collectively called 'reduction' in this study) is used to map high dimension vectors to lower dimensions in order to reduce computational cost and prevent data overfitting.

\begin{figure}
\centering
\includegraphics[width=6 cm]{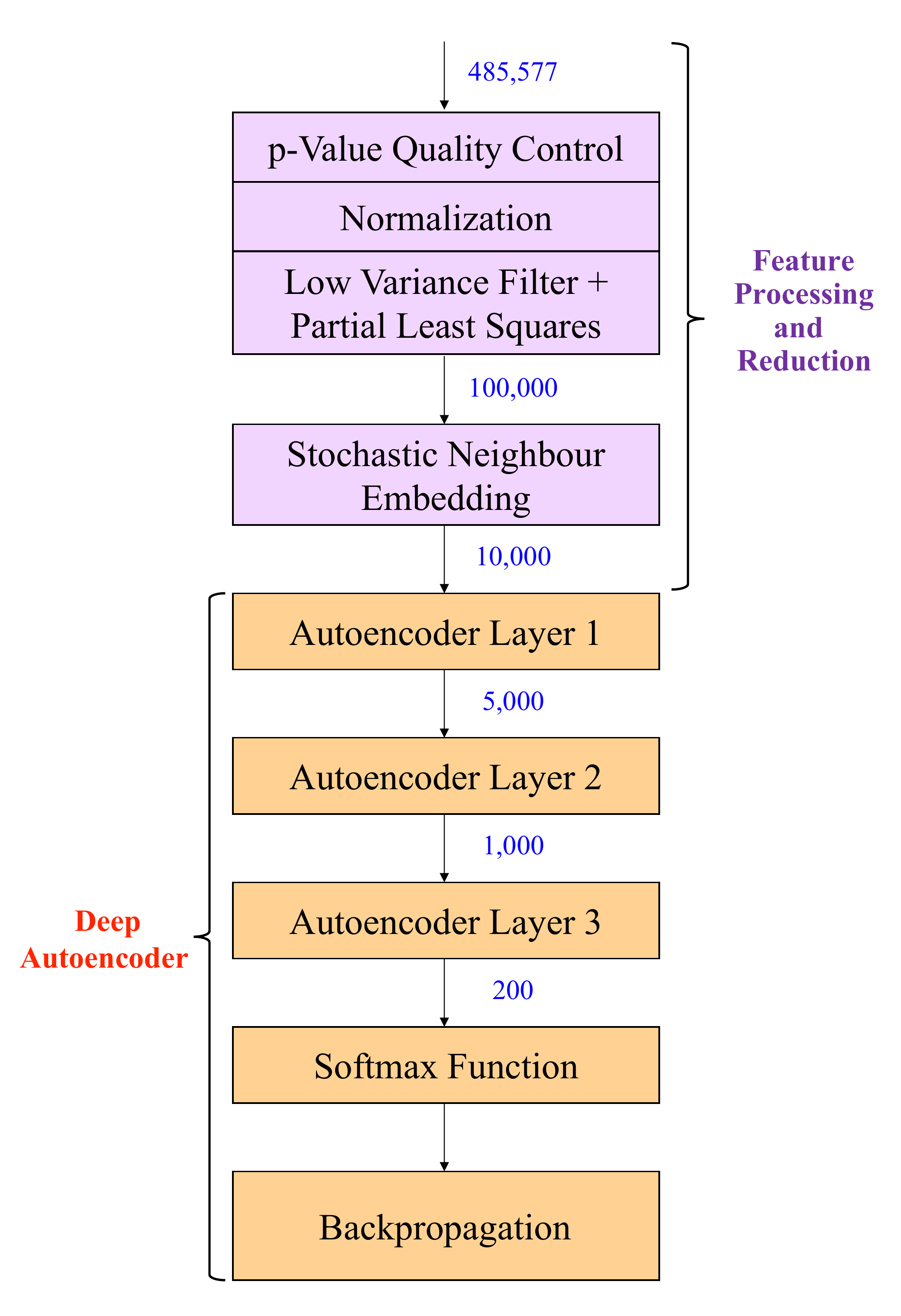}
\caption{Flowchart representing series of steps used in Feature Processing and Reduction, along with the Deep Autoencoder Network. The boxes and text in black represent the method used, while the numbers in blue represent the number of features extracted in each step.}
\label{fig:Learning_Disease}
\vspace{-1em}
\end{figure}

The feature processing and reduction flowchart can be seen in Fig.~\ref{fig:Learning_Disease}, illustrating the Deep Autoencoder network layers described in Section~\ref{sec:dae}. The first step involves a quality control of all the islands by obtaining $p$-values across all points, extracted by obtaining a ratio of the signal value for each probe (value) to the background signal obtained from the negative probe values, as illustrated in~\cite{maksimovic2016cross}. The $p$-value test is performed for all the island samples and a reliable feature set is then obtained. All signal values exceeding 0.01 are discarded while those below that value are kept as reliable quantities, before normalization of all the features is performed using the standard score given by:

\begin{equation}\label{eq:stan_score}
X_{Norm}=\frac{X-\mu}{\sigma}
\end{equation}

where $\mu$ and $\sigma$ represent the mean and standard deviation of each feature column. To scale, unity based normalisation is obtained from $X_{Norm}$ as:

\begin{equation}\label{eq:scaled}
X_{Scaled}=\frac{X_{Norm}-X_{Min}}{X_{Max}-X_{Min}}
\end{equation}

Reduction of features is then performed by using a low variance filter to eliminate all feature columns that possess low variance with respect to one another so that unwanted features that provide no useful information for differentiation between cancer classes are eliminated. This is then followed by the use of the Partial Least Squares (PLS) algorithm~\cite{boulesteix2006partial}, which in similar function as Principal Component Analysis (PCA), is used to compress the descriptor set to 100,000 features by creating a new feature space. PLS was used as it has been reported as a very efficient method for dimensionality reduction of genomic data, contrary to PCA, through the use of supervised covariance matrices~\cite{boulesteix2006partial}. Thus, the low variance filter eliminates features, while the PLS creates a new and reduced feature space. The reduced features are then clustered and reduced once again using Stochastic Neighbour Embedding (SNE)~\cite{hinton2003stochastic}, a well-known algorithm for reduction of high volume data-sets. The algorithm seeks to minimize the cost function $C$ of asymmetric probability $p_{ij}$ and induced probability $q_{ij}$, which were defined as by:

\begin{equation}\label{eq:sne_1}
p_{ij} = \frac{\exp(-d_{ij}^2)}{\sum_{k \neq i}\exp(-d_{ik}^2)}
\end{equation}

where $d_{ij}^2$ is the squared Euclidean norm of two high dimensional points $x_i$ and $x_j$.

\begin{equation}\label{eq:sne_2}
q_{ij} = \frac{\exp(-{||y_i - y_j||}^2)}{\sum_{k \neq i}\exp(-{||y_i - y_k||}^2)}
\end{equation}

where $y_i$ is a low dimensional object.

\begin{equation}\label{eq:sne_3}
C = \sum_1{\sum_j{p_{ij}\log{\frac{p_{ij}}{q_{ij}}}}}
\end{equation}

As described in~\cite{hinton2003stochastic}, the gradient can be simplified to:

\begin{equation}\label{eq:sne_4}
\frac{\partial C}{\partial y_i} = 2\sum_j(y_i - y_j)(p_{ij} - q_{ij} + p_{ji} - q_{ji})
\end{equation}

This helps significantly in forming clusters based on neighbours, thus combining and reducing the number of features from 100,000 to 10,000. Two different extraction algorithms (PLS and SNE) were used to reduce computational intensity and time, while producing the best features to be further utlized by the Autoencoder network. 

\subsection{Deep Autoencoder Network}
\label{sec:dae}

\begin{figure}[!t]
\centering
\includegraphics[width=\linewidth]{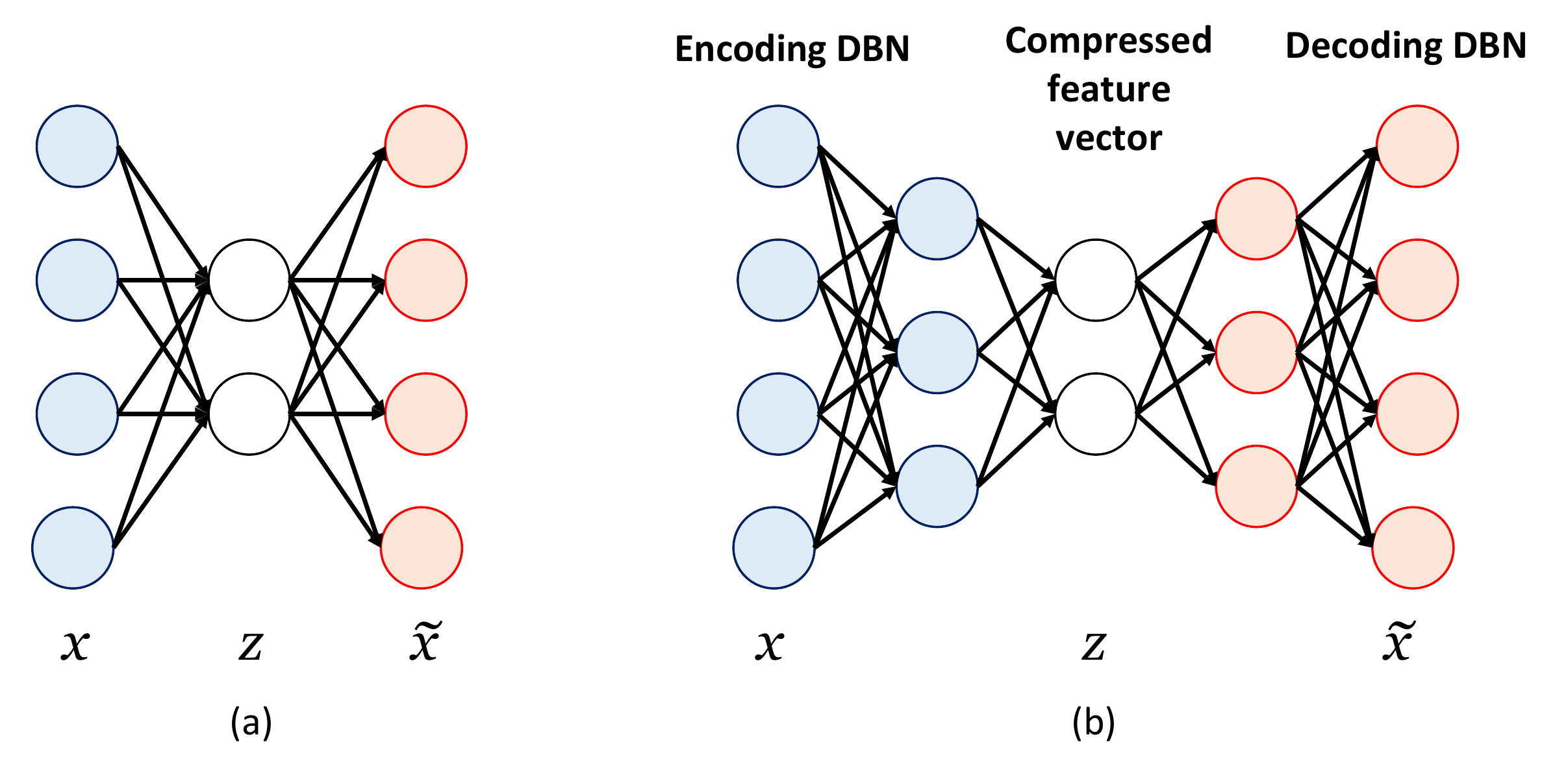}
\caption{(a) A basic Autoencoder consisting of an encoding and decoding layer; (b) Representation of a Deep Autoencoder consisting of two halves, each comprising a Deep Belief Network.}
\label{fig:dbn_da}
\vspace{-1em}
\end{figure}

The learning algorithm used in this research, the Deep Autoencoder, represents an unsupervised neural network that was created for the purpose of allowing efficient sparse representation of large data sets (encoding) and in time, the method has gained prominent significance as a learning tool. A basic Autoencoder design can be seen in Fig.~\ref{fig:dbn_da}(a) which forms a weighted structure similar to that of a Multilayer Perceptron, as it consists of two important halves, $encoding$, to compress the data into a sparse representation and $decoding$, to reconstruct the data to a form similar to the input.

Let $x_i$ represent the input data and $z_i$ represent its encoded form. As explained in~\cite{dl_auto_ref1}, the relation between them is given by:

\begin{equation}\label{eq:dae_1}
z_i = s(W_1x_i + b_1)
\end{equation}

where $W$ is the weight vector, $b$ is the bias and $s$ is the activation function. Essentially, $z$ is the encoded sparse representation of $x$. In the decoding half, $z$ is used to form the reconstructed output $\tilde{x_i}$, which is given by:

\begin{equation}\label{eq:dae_2}
\tilde{x_i} = s(W'_2z_i + b'_2)
\end{equation}

The sum of squared differences is the optimization function used. This is defined as:

\begin{equation}\label{eq:dae_3}
\begin{aligned}
J(W_1, W_2, b_1, b_2) & = \sum_i \Big(\tilde{x_i} - x_i\Big)^2\\
& = \sum_i \Big(s(W'_2z_i + b'_2) - x_i \Big)^2\\
& = \sum_i \Big(s(W'_2(s(W_1x_i + b_1)) + b'_2) - x_i \Big)^2
\end{aligned}
\end{equation}

To make sure that the weights and bias are optimized to produce $\tilde{x_i}$ that is very close to ${x_i}$, the sum function defined in Eq.~\ref{eq:dae_3} is minimized using Stochastic Gradient Descent. By using multiple layers of Autoencoders, a Deep Autoencoder network is created, consisting of two symmetric Deep Belief Network (DBN) layers, in which the first DBN half forms the encoding system, while the second DBN half forms the decoding system. A representation of such topology can be seen in Fig.~\ref{fig:dbn_da}(b), whereby each of the two DBNs consist of multiple constituent layers of RBMs for a Deep Autoencoder and the layer $z$ is used
 with a logistic or some other non-linear activation function to form the neural network classifier. 

In this research, the network consists of three layers mapping 10,000 descriptors in the first layer to 200 in the third. Together, these layers form an efficient neural network capable of CpG methylation feature classification and cancer type differentiation. Krizhevsky's method for image classification~\cite{krizhevsky2011using} served as the primary inspiration for incorporating Deep Autoencoders, as this work uses a Deep Autoencoder network to produce a content based classification of images from a multitude of classes and by using sparse 'codes' derived from large image features, a robust algorithm is produced. The unsupervised intermediate layers in the proposed network use a sigmoid activation function, while the supervised output layer uses a softmax function and classifies the cells into eight different categories. Learning rates between each iteration (epoch) are adapted, so that learning happens quickly in the first couple of epochs and fine tuning happens later, with Stochastic Gradient Descent to be used to minimise the error function. The neural network distinguishes between four cell types, and healthy/cancerous states, thus creating eight output classes to classify into.

\subsection{Technical Implementation}

All aspects of data cleaning, analysis, feature extraction (processing and reduction) are implemented using MATLAB and its associated toolboxes, with some algorithms being tested in Python, specifically the scikit-learn machine learning library~\cite{scikit-learn}. The Deep Autoencoder network is implemented using the Keras open source library~\cite{chollet2015keras} running on Theano. Experiments were run in a 16GB RAM CPU. 

\subsection{Performance Metric}

The performance of the system was evaluated based on the four criteria: Sensitivity ($SE$), Specificity ($SP$), Accuracy ($ACC$) and Matthew's Correlation Coefficient ($MCC$). These are calculated based on the accurate identification of positive (cancerous) or negative (healthy) cell lines. A True Positive (${TP}$) would indicate that the cancerous cell line is identified correctly, while a False Positive (${FP}$) indicates that a cancerous cell line is identified as healthy. Conversely, True Negatives (${TN}$) and False Negatives (${FN}$) are calculated for healthy cell lines. The metrics are defined by the following equations:

\begin{footnotesize}
\begin{equation}\label{eq:se}
SE=\frac{TP}{TP+FN}
\end{equation}
\begin{equation}\label{eq:sp}
SP=\frac{TN}{TN+FP}
\end{equation}
\begin{equation}\label{eq:acc}
ACC=\frac{TP+TN}{TP+TN+FP+FN}
\end{equation}
\begin{equation}\label{eq:mcc}
MCC=\frac{(TP*TN)-(FP*FN)}{\sqrt{(TP+FN)(TN+FP)(TP+FP)(TN+FN)}}
\end{equation}
\end{footnotesize}

While $ACC$ of the system is the most important performance indicator, the $MCC$ plays an equally important role in evaluating its performance as there are unequal numbers of healthy and cancerous cells. $SE$ and $SP$ determine the system's ability to separately identify healthy or cancerous cells, and gives a good indication of its ability. 

\section{Results and Discussion}
\label{sec:results}

An eight-category classification method is implemented to determine (i) whether a cell line is healthy or cancerous, and (ii) to distinguish the type of cell (and therefore cancer type and state) it associates with. From the demonstrated analysis, the achieved overall Sensitivity ($SE$) of the developed system was 88.24\%, followed by Specificity ($SP$) of 83.33\%, Accuracy ($ACC$) of 84.75\% and Matthews Correlation Coefficient ($MCC$) of 0.687. These values are higher for multi-cancer classification compared to previous state-of-the-art methods, as observed in Table~\ref{tab:comparison}, and the $ACC$ reported is for both the highest individual cancer class as well as for multi cancer type classification.\footnote{Multi cancer $ACC$ is outlined only for methods that classify more than one cancer types. It is more difficult to place samples into multiple classes when more than one cancer type is studied, as compared to classifying a single cancer type that has only two options for (mis)classification. For methods that classify a single cancer type, including~\cite{si2016learning}, their total $ACC$ is reported for Single $ACC$, while methods that use multi cancer classification report the highest individual $ACC$ for this metric.}

\begin{table}[!htb]
\caption{Achieved accuracy ($ACC$) and comparison with other published work}
\begin{center}
\resizebox{\columnwidth}{!}{
\begin{tabular}{cccc}
\toprule

\textbf{Method} & \textbf{No. of Cancer Types} & \textbf{Single ACC} & \textbf{Multi ACC}\\

\addlinespace
\midrule

\textit{Deep Autoencoder [This work]}  & 4 (8 classes) & 93.33\% & 84.75\% \\
\textit{RBM + SOM~\cite{si2016learning}}  & 1 (2 classes) & 97.06\% & - \\
\textit{Binomial Prob.~\cite{kang2017cancerlocator}}  & 3 (4 classes) & 83.5\% & 73.5\% \\
    
\addlinespace\bottomrule
\end{tabular}}
\label{tab:comparison}
\end{center}
\vspace{-1em}
\end{table}

\begin{figure}[!t]
\centering
\includegraphics[width=8.8cm]{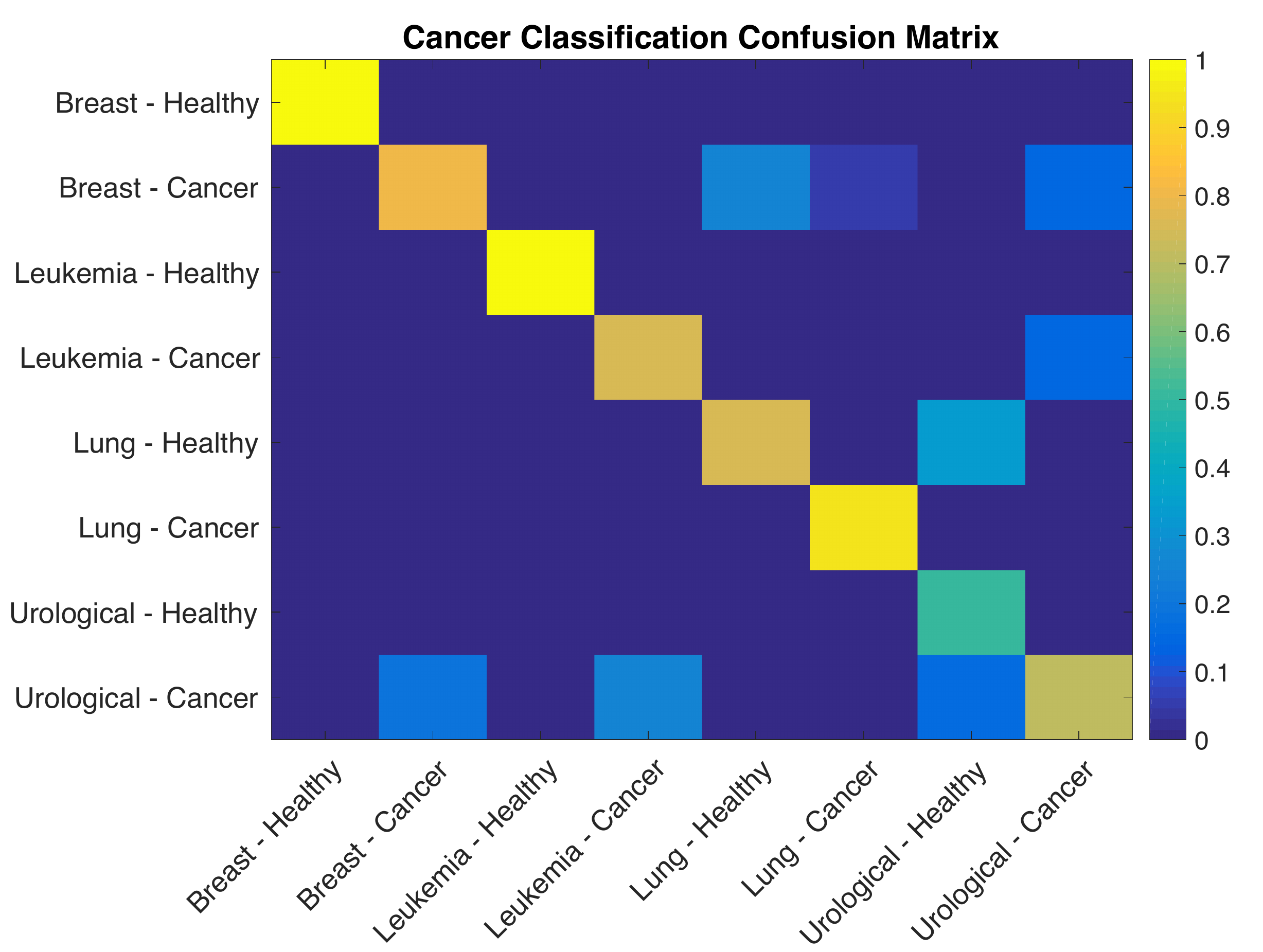}
\caption{Confusion matrix representing the accuracy of classification for each cancer type data-set. The higher the values of the diagonal elements (towards yellow), the more precise the classification. The figure illustrates that breast, leukemia and lung cell derived data-sets have little to no mistakes observed in classification, while urological cells (both healthy and cancerous) tend to be more often misclassified.}
\vspace{-0.7em}
\label{fig:cancer_conf}
\end{figure}

The results for the developed cancer classification method are plotted in a confusion matrix, shown in Fig.~\ref{fig:cancer_conf} and for each cancer type, the metrics obtained for different cancer types are reported in Table~\ref{tab:disease_metric}. From Fig.~\ref{fig:cancer_conf} and Table~\ref{tab:disease_metric}, it is observed that healthy cells of breast and white blood cells (leukemia) are classified very well, leading to 100\% specificity. In the case of lung samples, cancerous cells are classified very well, leading to a high sensitivity (94.73\%) and overall classification performance, white blood cells (leukemia) produce the best accuracy and MCC, indicating that prediction of cancer and its classification is very efficient. This was foreseen in Section~\ref{sec:obs} as leukemia cells possess clear differences in DNA methylation patterns, seen in their corresponding heatmaps. Breast and lung data-sets also perform fairly well with good accuracy and MCC values. Compared to other data sets, urological cell lines produced average scores as forecasted in Section~\ref{sec:obs}. This is attributed to the fact that healthy and cancerous cell lines for this type do not possess succinct properties in DNA methylation derived from the reported data-set used that can be used for good differentiation. 

\begin{table}[!htb]
\caption{Classification metrics of each cancer cell type}
\begin{center}
\begin{tabular}{c c c c c }
\toprule

\textbf{Cell Type} & \textbf{Sensitivity} & \textbf{Sensitivity} & \textbf{Accuracy} & \textbf{MCC}\\

\addlinespace
\midrule

\textit{Breast} & 80\%  & \textbf{100\%} & 87.50\% & 0.775 \\ 
    \textit{Leukemia} & 75\%  & \textbf{100\%} & \textbf{93.33\%} & \textbf{0.829} \\ 
    \textit{Lung} & \textbf{94.73\%} & 75\%  & 91.30\% & 0.697 \\ 
    \textit{Urological} & 71.42\% & 50\%  & 61.53\% & 0.219 \\ 
    
\addlinespace\bottomrule
\end{tabular}
\label{tab:disease_metric}
\end{center}
\vspace{-1em}
\end{table}

It is reported in~\cite{si2016learning} that for a single cancer cell (breast cancer cell) classification, the accuracy is 97.06\% and is very close to the results obtained in the case of leukemia and lung cancer obtained in the proposed method. The reason for higher accuracy in this research can be attributed to the fact that this method performs classification only for a single cell type (two classes) providing fewer opportunities for misclassification, while the proposed method distinguishes four cell types (eight classes). Nevertheless, for a multi-class system, the proposed method produces excellent results for individual cases, as compared to a single cancer classification system.

Whereas, in~\cite{kang2017cancerlocator}, it was reported that the overall accuracy obtained for four classes (three cancer types, and one non-cancer category) was 73.5\% with the presented method to produce an accuracy that is significantly better for four cancer types, with each healthy cell not being categorized into a single class, as proposed in~\cite{kang2017cancerlocator}, with the highest individual accuracy for a single class obtained for the non-cancer class, with an $ACC$ of 83.5\%, the proposed system achieves one of the highest reported individual $ACC$ of 93.33\% generated by the leukemia cell derived data-sets. The difference in error rates can be credited to the fact that the learning method described in~\cite{kang2017cancerlocator} uses a binomial probability distribution model, whereas the proposed method uses multiple probability models (deep learning layers) to better identify differences between classes. 

The proposed method also offers the novel distinction of using DNA methylation prediction to estimate values of missing islands, which would have otherwise been ignored, preserving more information from partial data while providing a better overall accuracy. While the goal of this research is the development of a strong machine learning algorithm for classification and differentiation of cancer types, which has been successfully demonstrated; it is also inferred that access to a larger and unified DNA methylation database would have improved the system's ability to classify some individual cases, as in the example of urological tumors, which would be within the scope of a future investigation. 

\section{Conclusion} 
\label{sec:conclusion}

A deep learning based cancer prediction and classification system has been described in this paper, using a prediction assembly to provide comprehensive information for data-sets derived by cancerous/healthy cell lines. Forming a correlation with cancer specific biomarkers and their prevalence in gene-specific and genome-wide genomic regions, patterns of DNA methylation were used as indicators. In conjugation with a combination of feature processing and dimensionality reduction methods, coupled with a Deep Autoencoder network, the proposed system produces a robust machine learning arrangement, capable of classifying four given cancer types. The intricate, yet well-defined pipeline of state-of-the-art-methods applied to complete CpG island features of human cell lines allows the system to select the best features, producing an overall accuracy of 84.75\% to distinguish four cancer types and is significantly better than the closest work in literature reported to classify three cancer types.

The proposed work is a Minimum Viable Product (MVP) that provides the required functionality to distinguish cancer types, offering a visual representation of the generated classification outcome. The developed method is envisioned to be further adapted in the artificial intelligence based systems that would provide decision support to healthcare professionals, offering methods for improving patients stratification and diagnosis. In conjugation with cancer diagnostic and prediction systems, computational models will thus allow for the creation of a unified big data driven platform to aid and personalise cancer diagnosis and predict response of patients to treatment, offering the clinicians further insight, allowing patients diagnosed with cancer to undertake treatment in a faster and more effective way. This work will contribute towards this envisioned future, improving cancer care and progressing the future of technologies in cancer research.

\section*{Acknowledgement}

The authors would like to acknowledge EPSRC (EP/N002474/1) and ERC (319818/i2MOVE) for supporting this research.

\ifCLASSOPTIONcaptionsoff
  \newpage
\fi

\bibliographystyle{ieeetr}
\bibliography{References}
\addcontentsline{toc}{section}{References}

\end{document}